\documentclass[conference]{IEEEtran}
\IEEEoverridecommandlockouts
\usepackage{cite}
\usepackage{amsmath,amssymb,amsfonts}
\usepackage{algorithmic}
\usepackage{graphicx}
\usepackage{textcomp}
\usepackage{xcolor}
\def\BibTeX{{\rm B\kern-.05em{\sc i\kern-.025em b}\kern-.08em
    T\kern-.1667em\lower.7ex\hbox{E}\kern-.125emX}}
\begin{document}

\title{Physical Design: Methodologies and Developments
}

\author{\IEEEauthorblockN{Abhay Chopde\IEEEauthorrefmark{1},~\IEEEmembership{Fellow,~IEEE}
 and
Atharva M. Kulkarni\IEEEauthorrefmark{1},}
\IEEEauthorblockA{\IEEEauthorrefmark{1}Dept. of Electronics and Telecommunication Engineering,
Vishwakarma Institute of Technology, Pune, India}

}

\maketitle

\begin{abstract}
The design and production of VLSI chips is a multilevel heirarchical process. As the demand for reduced die-area and technology nodes becomes prevalent, it gets increasingly challenging to optimize Power, Performance and Area (PPA) parameters to accommodate for the ever-increasing core logic on a chip. A well defined heirarchical flow is thus quintessential when it comes to VLSI design process. A robust heirarchical flow should encompass all stages, right from Gate-level RTL Synthesis (Front End Design) to Logic Placement and Verification (Back End Physical Design) and finally culminating with tapeout / production. Physical Design in this aforementioned flow is the process of translating logical circuit description into physically realizable GDSII form. This involves defining the best possible placement and routing for standard cells, macros and I/Os in the design to optimize PPA for any given netlist. This paper helps capture the nitty-gritty of methodologies and algorithms that are pertinent to the building and optimization of an efficient and robust physical design flow in VLSI chip-designing process.
\end{abstract}


\section{Introduction}
Very-large-scale Integration (VLSI) technology revolves around designing Generic or Application Specific Integrated Circuits (ASICs) with the core logic sometimes comprising of billions and even trillions of transistors, all embedded within a single small chip. Monumental developments in the field of VLSI date back to the 1970s when we were experiencing a new dawn in semiconductor physics and its application in building System-on-chip (SoCs), faster Processors and communication technologies.[1,2] Prior to all these technological enhancements, most ICs and processors could perform just a handful of operations with limited logic and instruction sets. With the help of developments in VLSI field however, it has become possible to implement circuits displaying desired levels of performance parameters based on industry requirements. Such designs may contain billions of standard and physical cells, I/O pins, macros, analog IPs, PLLs, data generators and numerous other blocks. The trend in VLSI technologies was observed and documented by Gordon E. Moore as early as 1965. The famed Moore's law states "The number of transistors in a microchip doubles every two years, though the cost of computers is halved".[1,2] However, we now have entered an era wherein this law may no longer hold validity owing to the complexities involved while working at lower process nodes. Despite these complexities, the cutting-edge EDA tools developed by Cadence Design Systems® and Synopsys have been aiding Design Engineers to produce chips showing desired characteristics, applications, processing parameters and portability. 

Physical Design is the process of translating the gate-level RTL logical functionality of a design (.vg) into a physical geometricized form (GDSII) which can be taped-out for production / packaging.

Need for Physical Design:

\begin{itemize}
    \item Current IC designs have millions of transistors and other complex logic which is routed within each other along with several layers of metal in a given metal stack.
    \item Such designs furthermore need to be optimized for Power, Performance, Timing and Area so as to ensure the best possible performance output with lesser setbacks.
    \item Manually optimizing the Placement and Routing of all components in a design has become a Herculean task being time consuming and error-prone. As a result, Automation in EDA industry is quintessential to carry out such an arduous task.
    \item Thus, a well-defined robust PD flow enhances our time to market capability and get more work done.
\end{itemize}

Physical Design in itself is a complex multi-domain process. As a result, it is broken down into simpler sequential heirarchies to expedite the design process:

Partitioning	breaks up a complex top-level circuit into smaller blocks or modules which can each be designed, optimized and analyzed in isolation, before merging them back into top-level analysis.
Floor Planning	determines the block dimensions, upsizing of HIC cells; assignment of pins, boundary and well-tap cells along with preplacement of any critical logic in the design. 

Power Planning	often done in conjunction with floor planning, distributes power rings, rails and subsequent VDD, VSS, VDDQ, VDDR etc power domains across the design. Power stripes are added in accordance with the metal stack approved by the industry.
Placement	is done following the floor planning stage, where the tool automates the placement of core logic, macros and decap cells within the die area.
Clock Tree Synthesis	involves building and routing clock architecture throughout the design. Critical parameters like Root pin, Through pin, Unsync pin and splits need to be defined prior to this stage, to balance skews and slack timings in the clock path.
Routing	is done to define metal fill and connectivity between all the logic implemented in a design. Metal layers and corresponding vias are dropped into the placed logic in accordance with the metal stack and sub-circuit requirement.
Timing Closure	helps optimize timing measures like clock skew, delay skew, max- trans, setup and hold slack in clock as well as data paths. Timing closure is critical as failure to meet timing requirements can cause corruption in data or on-chip instruction sets.

Verification stages include:

\begin{itemize}
    \item Static Timing Analysis (STA)
    \begin{itemize}
    \item Setup and hold check
    \item DRVs (max tran, max cap, fanout) check
    \item Clock and Data skew check
\end{itemize}
    \item Power Distribution Network Analysis (PDN)
    \item Physical Verification (PV)
        \begin{itemize}
    \item SI noise and Antenna check
    \item Design Rule Checks (DRCs)
    \item Layout vs Schematic (LVS)
    \item DFM, LPA Analysis
\end{itemize}
    \item Formal Verification (FV)
    \item Low Power Verification (CLP)
\end{itemize}

\section{LITERATURE SURVEY}
Application Specific Integrated Circuits (ASIC) design, as the name suggests tends to cater to specific design purpose sought out by the industry. Such specific purposes may be pertaining to chip performance, clocking frequency, data requirement or process node, to name a few. The term "Technology Node" or "Process Node" is held in high regard in the semiconductor industry. Process node denotes the size of the smallest gate being implemented in the chip. Smaller the process node, smaller will be the gate size, greater will be the gate density per micron, thereby resulting in greater scope to implement logic onto the chip. For example, a 7nm process node itself can have approximately one-tenth billion transistors per sq. mm.[3,4] In 2021, Intel has revealed its plans to launch technologies based on 2nm process node as early as in the year 2024.[5] Although, greater gate density gives more power to designers to implement logic, it complicates the process of design furthermore leading to difficulties in closing timing and optimizing PPA. Lower process nodes (especially below 32nm) inherently add a layer of complexity especially in terms of DRCs and congestion hotspots. Inamul Hussain and Saurabh Chaudhury published some research in 2020 discussing the prominence of power dissipation and leakage problems at such process nodes. Selection of appropriate logic family is therefore deemed critical at the start of designing any functional chip. As far as static logic families go, their research claims that CNTFETs (Carbon Nanotube Field-Effect Transistors) surpass MOSFETs (Metal Oxide Semiconductor Field-Effect Transistors) owing to their lower static power and leakage levels at lower technology nodes.[6]

Developments in the EDA industry have been critical in overcoming these challenges. With hi-tech EDA tools offered by prominent benefactors to the silicon industry in Cadence Design Systems and Synopsys Inc, it is possible to optimize results for majority of the designs and logic families, even MOSFETs. These tools are compatible with representing data from all major foundries including Samsung, TSMC, Intel etc. ; reading in multiple design inputs like timing libraries, derate values, LEFs and metal stacks ; synthesizing the apt netlist and constraints to serve chip functionality ; carrying out placement and routing of all cells and logic in the design ; implementing verification mechanics for timing parameters, LVS, DRCs etc.[7]. Recently, Cadence Design Systems has even been able to integrate Artificial Intelligence into chip-designing with the development of its "Cerebrus" AI-oriented EDA solution. Rest assured, the EDA industry is always on the move to provide better and more robust solutions to the challenges faced by the ever-changing silicon market.

The solutions offered by EDA tools are gauged with the aid of several industry metrics to ensure accurate outputs and eradication of bottlenecks when the chip is out for silicon testing, fabrication and packaging. Considering the ever-increasing demands in the near future in terms of processing power, supercomputing combined with the desire to implement greater logic on a smaller area, Aaron NG and Igor Markov have proposed a robust benchmark mechanic to test the performance of EDA tools. Such a benchmark helps gauge the tool in terms of accuracy, possibilities of failure, computing capacity, ability to extrapolate graphical and tabular results, Pareto Regression Analysis to compare Quality of Results (QoR) with similar tools or previous versions, ability to control the instruction flow etc.[8]

Cutting-edge EDA tools provide us with high level design automation to help meet design requirements and constraints with minimal manual work. However, it is quintessential that we provide these tools with robust design methodologies and flows so as to get the best possible results. Three design architectures were predominantly in use in the early 1970s all the way upto the 90s to propose semiconductor designs for embedded CMOS applications, namely ASIC, DSP and RISC.[9] However, since the dawn of this century, owing to the heaps of progress made in the silicon frame pertaining to chip size and on-chip computation, ASIC architecture has been at the heart of nearly all proposed VLSI designs. A standard ASIC flow covers all steps right from RTL synthesis to optimal Physical Design and Verification to finally culminating with production (tapeout). All stages in this process are critical. Failure to meet design constraints in any of these stages may cause hiccups like data corruption, false outputs or even worse, chip failure. As a result, it is equally critical to maintain a well defined flow which ensures that performance during all the aforementioned stages is met as per desired requirements. Such ASIC flows can be implemented on FPGAs or CPLDs depending upon the dimensions and magnitude of proportions of the design along with functionality and re-programmability. FPGAs generally supprt larger designs than CPLDs as they can host a large number of sequential flops and registers. Being RAM-based or volatile in nature, FPGAs offer more flexibility at the cost of a slightly higher booting or configuration time. Once configured however, FPGAs are capable of fast-computing on-chip signals. CPLDs being ROM-based however are able to load up memories immediately on booting up, but are slower in computational capability in the longer run.[10,11] Traian Tulbure has published some of his work on the reconfigurable nature of CPLD logic and its viability in ASIC implementation in 2011. His research talks about the dynamic reconfigurable nature of CPLDs and how it could have an edge over SRAM-based implementation for smaller designs. Satisfactory results were obtained from a timing point of view over the course of this research.[10]

In the ASIC Design flow discussed earlier, Physical Design is the process following Gate-level RTL synthesis and involves representation of gates defined in the synthesis verilog netlists into their geometricized forms complete with physical connectivity provided using appropriate metal layers. This geometric representation is in GDSII form which is easily realizable for masking and tape-out during production stage. As we look deeper into the Physical Design flow and all the methodologies that accompany it, we realize that it follows a sequential heirarchy consisting of several sub-stages involving layout, timing and verification of the layout.
The new age IC design flow consists of three most critical sub-sections.
        \begin{itemize}
    \item RTL Synthesis (Front End Design)
    \begin{itemize}
    \item This step involves synthesizing a verilog netlist and constraints in Hardware Description Language (HDL) and Synthesis EDA tools in accordance with the die area, process node, performance, chip functionality and metal stack amongst numerous other requirements set by the Foundry.
\end{itemize}
    \item Physical Design (Back End Design)
    \begin{itemize}
    \item As discussed earlier, Physical Design takes the Synthesis .vg and constraints as input and translates it to physical geometries to be implemented on the chip. Throughout this process, the goal is to optimize the design from timing and PPA point-of-view.
\end{itemize}    
    \item Physical Verification
    \begin{itemize}
    \item The objective of Physical Verification is to check the design for any opens, shorts, unconnected pins and Design Rule checks set by the foundry as well as validating the physical layout of our design against the schematic obtained from the foundry.  
\end{itemize} 
\end{itemize}
After the design is validated from verification point of view, it is then sent out for Tape-out and fabrication, followed by packaging. This post-verification process involving masking and chip production is handled by the Wafer Fabrication Houses before being packaged, tested and implemented as ICs. 
All these stages in ASIC design flow possess their own different methodologies and architecture which offers ease of interface for EDA tools as well as better conductivity of tasks. Along with the verilog netlists (.vg) and constraints generated during Synthesis stage, there is some additional foundry data which needs to be provided as input to the Physical Design stage. This data is commonly known as "foundry collaterals" or "process collaterals" and contains information regarding attributes such as leakage power, area and functionality of standard cells with respect to different PVT corners, abstract view for the design and definition of metal layers.

Cortadella et al. have published a journal article in 2015 taking us through the steps from combinational and sequential logic synthesis leading upto automated pipelining options for High-level Synthesis (HLS). Execution time for any synthesis algorithm is governed by the following equation:
\begin{figure}[htbp]
\centerline{\includegraphics[height=70pt,width=0.3\textwidth]{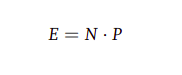}}
\caption{Performance Equation for Synthesis Algorithms}
\label{fig}
\end{figure}
where N refers to number of cycles required to carry out instruction set and P denotes the time period for each of the N cycles. In combinational logic synthesis, instructions are provided to try optimizing critical paths by working around the value of P and other combinational logic. Sequential logic synthesis deals with optimizing false and multicycle paths in the design and tampering with flops and other sequential components in the design.[12] Research work done by Satoshi Ohtake et al. further elaborates on accurately identifying false paths in a RTL design. They have proposed a novel methodology titled 'Mapping Point Preserving-Logic Synthesis (MPP-LS)' which maps path-to-path logical connectivity between the sequential logic in a circuit and distinguishes actual logical paths from redundant ones.[13]
Front End Design further encompasses the following design processes:

Design entry: Each chip or IC is designed for a specific purpose set by the industry. This purpose is defined in the form of implementation logic on the chip. The objective of this stage is to identify this core logic and process settings along with the list of collaterals and desired requirements specified by the industry. Foundry collaterals for this stage typically include architectural recommendations, target frequency for the design, timing windows and corresponding waveforms, MMMC corners etc. 

Logic Synthesis: The functionality defined in the previous stage is put into code during Logic Synthesis. This code is written in any Hardware Description Language (HDL) such as Verilog or VHDL. This RTL logic is passed over to EDA Synthesis tools such as Genus™ by Cadence® to obtain the Gate-level netlist as output.

Gate level Simulation: This is a post-synthesis validation procedure to verify the functionality of generated verilog with industry expectations. This stage also involves generation and analysis of power, timing and density reports for the synthesized netlist and corresponding constraints.

Authors A. Kahng, J. Lienig et al. in their 2011 Springer publication titled "VLSI Physical Design: From Graph Partitioning to Timing Closure" have garnered a comprehensive research base on Back End VLSI Flow. Their work encompasses detailed study on the following Physical Design methodologies.
The back-end design includes following steps.[14]

Schematic Entry: Similar to the Design Entry stage in Front End flow, this stage involves reading-in the logic design needed.

Pre Layout Simulation: The logic design extracted in the previous step is validated before moving forward to the layout stage. The simulation involves verification of the synthesized netlist with schematic provided by the foundry with the help of Cadence® Ultraism EDA tool.

Design Layout: After the logical equivalence of synthesized netlist is verified, the design is ready to enter layout stage. The Innovus™ and IC Compiler II EDA tools developed by Cadence® and Synopsys respectively are widely used for layout and subsequent optimization. The layout stage pans the following sequential methodologies:

\begin{itemize}
    \item Floor Planning
    \item Power Planning
    \item Placement
    \item Clock Tree Synthesis
    \item Routing
    \item Timing Optimization
\end{itemize}

Extracted Simulation: The design obtained from first-cut P\&R needs to be optimized for power, performance, timing and area while also being wary of the timing and physical violations present in the design. Quantus™ which is an extraction engine developed by Cadence® aids in extraction of resistive and capacitive parasitics arising due to interconnect wires added to the layout while routing. This net delay information is stored in Standard Parasitic Exchange Format (SPEF) files and is further used for timing analysis. While clearing violating paths occurring in timing report, it is typical for designers to try and optimize cell and net delay values in the layout.

The Journal "Fundamentals of Layout Design for Electronic Circuits" (Springer 2020) authored by J. Lienig and J. Scheible talks in depth about various library interfaces, design rule checks and resulting violations. Extensive research that has been done into different EDA tools along with different heirarchies and flows followed by each tool is another highlight from this journal.[15]

\section{BACK END FLOW}
\begin{figure}[htbp]
\centerline{\includegraphics[height=220pt,width=0.5\textwidth]{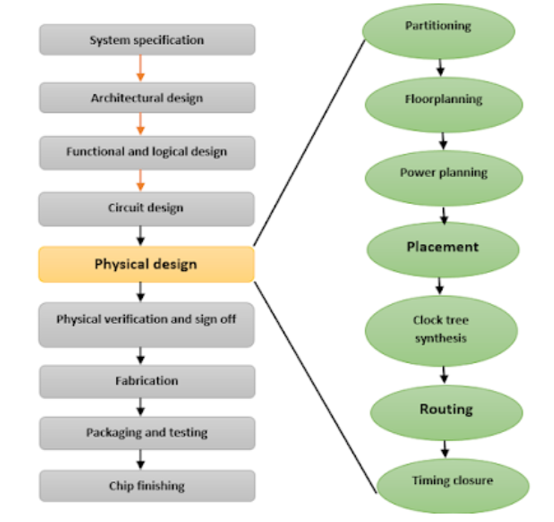}}
\caption{ASIC Design Flow}
\label{fig}
\end{figure}

\subsection{Physical Design}

A 2021 research by Dmitry Bulakh et al. published at the International Seminar on Electron Devices Design and Production was aimed at developing a Graphical User Interface (GUI) to visualize, control and interpret the various stages of Physical Design. The framework was developed in C++ using a few inherited classes from a cross-platform Qt library. Some .hpp, .dll and lib files can been provided as input along with layer map files, and the framework would then render a GDSII file as output.[17]

The main steps in Physical Design flow are:
\begin{itemize}
    \item Post-synthesis Netlist Optimization
    \item Floor Planning
    \item Power Planning
    \item Placement (Pre-CTS)
    \item Clock Tree Synthesis (CTS)
    \item Routing
    \item Post Route Timing Optimization
    \item Physical Verification
\end{itemize}
. 
\subsection{Design Collaterals}

\begin{table}[h]
\begin{tabular}{ll}
Gate Level Netlist (.vg)                                                            & \begin{tabular}[c]{@{}l@{}}This is the output file generated \\ after synthesis. It is the gate level \\ representation of the design.\end{tabular}                                                                                                                                                                                                                                                                                                                                                                                                                                                                                                                                                                                                           \\
Technology file (.tech)                                                             & \begin{tabular}[c]{@{}l@{}}\\This file is provided by foundry /\\ fabrication team. It provides technology \\ specific information like physical and \\ electrical characteristics of metal layers, \\ vias and metal widths, spacing, pitch \\ and routing design rules\end{tabular}                                                                                                                                                                                                                                                                                                                                                                                                                                                                           \\
\begin{tabular}[c]{@{}l@{}}Logical libraries / \\ Liberty files (.lib)\end{tabular} & \begin{tabular}[c]{@{}l@{}}\\This file is provided by process foundry. \\ It contains information regarding PVT \\ requirements, net delays, cell delays, \\ transition, recovery, removal, setup and \\ hold time requirements. It also contains \\ information about area of cell, leakage \\ power, capacitance etc. LIB files are \\ generated using either of Composite Current \\ Source (CCS) or Non-linear Delay Model \\ (NLDM) or ECSM methodologies. For smaller \\ technology nodes, CCS is being preferred.\\ The design needs to be validated for certain \\ PVT (Process, Voltage and Temperature) \\ corners to ensure seamless functionality \\ under even the harshest of conditions. \\Timing is different for different analysis \\ views and corners. Hence, there is a .lib file \\ for every PVT corner.\end{tabular}
\\
\begin{tabular}[c]{@{}l@{}}Library Exchange \\ Format (.lef)\end{tabular}           & \begin{tabular}[c]{@{}l@{}}\\This is provided by the foundry itself. \\ The LEF is an abstract view of the cells\\ It contains information about cell geometries, \\ routing and via placements.\end{tabular}                                                                                                                                                                                                                            \\
\end{tabular}
\end{table}

\begin{table}[h]
\begin{tabular}{ll}
\begin{tabular}[c]{@{}l@{}}System Design \\ Constraints file \\ (.sdc)\end{tabular} & \begin{tabular}[c]{@{}l@{}}\\Constraint files are generated during synthesis \\ phase in accordance with foundry requirements \\ pertaining to timing, power, performance and area \\ requirements of the design. These files also define \\ the following: operating conditions, DRVs (max \\ trans, fanout and capacitance), frequencies of \\ source and generated clocks along with clock \\ uncertainty and latency, multicycle and false paths \\ etc. along with numerous other constraints.\end{tabular}  
\\
\begin{tabular}[c]{@{}l@{}}Table Look Up \\ (TLU)\end{tabular}                      & \begin{tabular}[c]{@{}l@{}}\\TLU is a binary file used for RC estimation and \\ extraction, although header is in ASCII format. TLU \\ contains wire capacitance at different spacing and \\ width in the form of a look-up table which provides \\ high accuracy and runtime benefits. \\ This file provides RC parasitic of metals per unit \\ length which is subsequently used to calculate net \\ delay.\end{tabular}
\end{tabular}
\end{table}
. \\ \\ \\
\subsection{Floor Planning}
As the name suggests Floor Planning helps create a skeletal level framework for spatial locations of standard cells, macros, analog IPs and all other blocks on a circuit. The goal is to optimize design layout on the given die area while keeping close tabs on probable congestion and density violations. Typically floor planning is done to make the layout compact wherein logically connected instances are placed in close proximity to each other as well as everything else. This is done to make effective use of the routing resources available.

\begin{figure}[htbp]
\centerline{\includegraphics[width=0.5\textwidth]{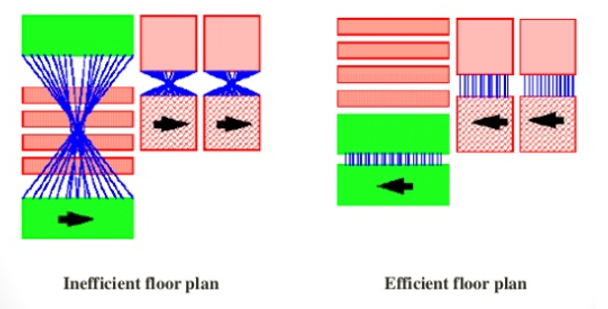}}
\caption{Example of efficient floor plan}
\label{fig}
\end{figure}
Goals for Floor Planning:
\begin{itemize}
    \item Minimize the total chip area and dead space
    \item Minimize total wire length
    \item Minimize Interconnection complexity
    \item Improve the performance by minimize delay
\end{itemize}

Naushad Manzoor Laskar et al. have documented a comprehensive study of all the prominent Floorplan Representations and the means of achieving them in their 2015 publication titled "A Survey on VLSI Floorplanning: Its Representation and Modern Approaches of Optimization". This detailed research covers all aspects of floorplan and floorplanning algorithms starting with all different ways of denoting a floorplan.[18] Different ways of representing a floor plan are as follows:
\begin{itemize}
    \item Polish\\Polish notation is widely used in the world of computing to read and express logic embedded in data structures, but most prominently binary trees in the form of equations. Such a notation is used to read the information conveyed by the sliced up floorplan binary tree in post-fix form.
    \begin{figure}[htbp]
\centerline{\includegraphics[width=0.4\textwidth]{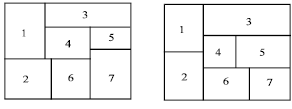}}
\caption{Non-slicing v Slicing Floor Plan}
\label{fig}
\end{figure}
    \item Sequence Pair
    \item Bounded Slicing Grid (BSG)
    \item TCG
    \item O-Tree\\The O-Tree algorithm is one of the least computationally expensive algorithms with a complexity of O(n). It is a local-search algorithms and hence deterministic by nature. Being deterministic means that this algorithm will try to optimize the layout on the basis of immediate short term solutions in contrast to a few greedy algorithms which are a bit more experimental in their approach and tend to be better at optimizing. Despite being one of the most computationally efficient algorithm, O-Tree based approach may not always find the best possible solution owing to its deterministic nature. A research done in 2005 by Maolin Tang and Alvin Sebastian addresses this very issue and proposes a more greedy Genetic Algorithm (GA) to optimize Floor Plan based on O-Tree representation.[19]
    \item B* Tree\\Similar to O-Tree, the B* algorithm showcases a computational complexity of O(n). It is widely regarded as one of the most efficient and flexible Floor Plan notations to optimize upon. It makes use of ordered weighted binary tree, the root of which is located at the bottom left corner of the placement area at the coordinate (0,0).Once the root is fixed, the rest of the tree is built recursively, first populating the left branch and then the right.
    
\end{itemize}
.\\
Optimization\\
Once the initial Floorplan notation is decided upon, then begins the process of optimizing the layout to have the smallest area with the most optimum core utilization. [18]
\begin{itemize}
    \item Simulated Annealing\\Simulated Annealing is the oldest Floorplanning algorithm and has been extensively used over the years. It can be used effectively with slicing as well as compacted notations like Polish, B*Tree etc. This algorithm can converge to a fairly optimal solution but is irregular in doing so. In modern day, this algorithm is used as a preliminary benchmark for researchers to test newer algorithms on.[18]
    \item Genetic Algorithm\\Soon after the success of Simulated Annealing algorithm, the Genetic Algorithm (GA) was developed. GA optimization begins with arbitrary placement of blocks on the pre-defined die area. The Cost Function of this initial floorplan is calculated using some distance metrics. Then any two of the blocks are spatially swapped and a new revised cost function is deduced. This is then compared with the initial value and a decision is made whether the swapping has affected the layout positively or adversely. This process is recursively carried out until the most optimum value is obtained for the cost function.[18]
    \item Partical Swarm Optimization (PSO)\\In PSO, each block is treated as an individual entity with the same weight. The premise of this algorithm is to spatially change the position of these individual blocks so as each block determines its own best position with respect to its nearest neighbours in the design. A research published in 2019 by S.B.Vinay Kumar et al. proposed an Adaptive PSO model which tunes the PSO model further by adding a weight factor to the block. All blocks are defined weight values at the initial stage. Over successive iterations as the block gets closer and closer to determining its optimum posititon, its weight and hence priority for optimization keeps on reducing. So all blocks will start with a high inertia value and end at a smaller value towards the completion of the optimization process. Such an adaptive architecture is more likely to give better results than GA and traditional PSO models, as claimed by the study.[20]
    
\end{itemize}
As all the above discussed algorithms point out, optimization is a recursive process and hence may take up a lot of time to converge on an optimal floorplan. To reduce the time taken by traditional floorplan algorithms, Yanling Zhou et al. have proposed a quicker floorplanning method which involves breaking the full initial netlist into smaller parts and carrying out floorplanning for each individual sub-netlist in a parallel threaded manner thereby saving time on converging to the optimal floorplan.[21]

As discussed earlier, floor plan lays out a framework for the physical design to be built upon. While the process of defining a floor plan may be entirely different for designs with different level of heirarchies, there are a few common fields to be defined to ensure an efficient design layout. To begin with, floor plan defines the block dimensions thereby setting the die area. To make optimum use of this die area, it is critical to identify all the logic, IPs, macros, I/Os and memories which need to be placed in the design. For placement of all the aforementioned components, floor plan should take into account the possibility of cell density issues or routing congestion that may arise owing to poor or incorrect placement. The designer should also aim to make the layout compact so as to use routing resources efficiently as well as to avoid wastage of die area. Core area utilization should also be taken into account while define floor plan. Most industries target 60-70\% initial core utilization to generate margin for timing optimization at a later stage. A compact floor plan presents a few advantages related to speed of operation of the chip, reason being, the more compact our design, closer will be the placement of on-chip components, lesser will be the routing resources used, lesser will be the interconnect length and subsequent net delay, thereby reducing net latency and increasing the speed of operations. This gives rise to an active trade-off between speed of design and resulting routing congestion.
 
Common steps panning the Floor Plan stage include:

\begin{itemize}
    \item Partitioning
    \item Defining Block Dimensions
    \item Pin Placement
    \item Adding Decap, Tap and End Cap cells
    \begin{figure}[htbp]
\centerline{\includegraphics[width=0.5\textwidth]{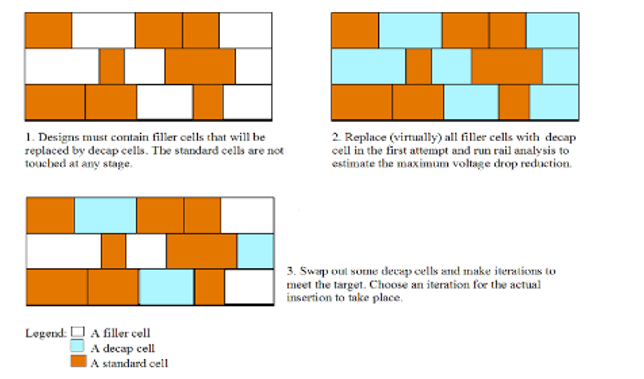}}
\caption{Adding Decap cells}
\label{fig}
\end{figure}
    \item Macro Placement
    \item Adding Routing and Placement blockages
    \item Adding IO buffers
    
\end{itemize}

\subsection{Power Planning}
Power Planning is the process involving power grid creation to facilitate equal distribution of energy to all parts of the design.

What creates the Power Grid?
There are 3 levels of power distribution involved:

1.	Rings: Carries VDD and VSS around the chip

2.	Stripes: Carry VDD and VSS from the Rings around the chip

3.	Rails: Connect VDD and VSS from chip level to standard cell level

    \begin{figure}[htbp]
\centerline{\includegraphics[width=0.5\textwidth]{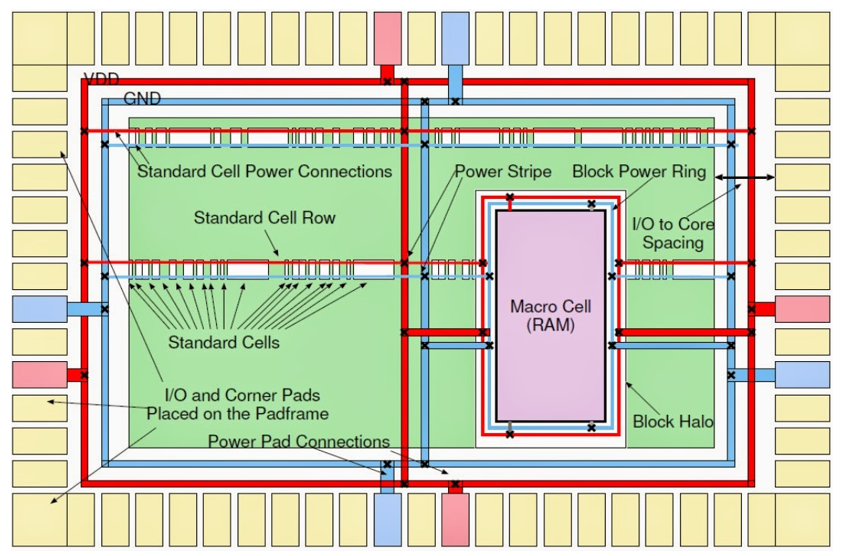}}
\caption{Power Planning}
\label{fig}
\end{figure}

Steps involved in this stage:
\begin{itemize}
    \item Width, pitch and offset dimensions of power stripes wrt each metal layer in accordance with provided metal stack.
    \item Block and I/O Power connection at top level using power rings, bumps and stripes.
    \item PG connection at standard cell and block level via power rails.
    
\end{itemize}

Power Planning is a very critical stage in Physical Design as it defines the Power/Ground mesh for any layout. The mesh needs to be built considering power requirement for all pins and macros in the design ; standard cell placement and the possibilty of congestion or high IR drop must also be taken into consideration while doing power and bump planning for any design. Any high voltage drop in the layout caused by EM/IR imbalance may cause circuital failure. Zhu Qing et al. in their publication titled "Simulation and Planning Method for On-Chip Power Distribution – An Industry Perspective" have studied the different parameters affecting power distribution and have come up with a set of steps to be followed while power planning to minimize the chances of fatal voltage drop across the design.[22]
The proposed steps are as follows:
\begin{itemize}
    \item Categorize all standard cells in the design into one of the two groups: those giving rise to severe IR drop and those with nominal IR drop. 
    \item Based on the IR drop metrics obtained in the previous step, draw a schematic for the power grid, with the horizontal and vertical layers being represented by a metal resistor. Compute the total RC parasitic offered by this schematic grid arrangement. For better accuracy, we may even include via parasitics in the calculation.
        \begin{figure}[htbp]
\centerline{\includegraphics[width=0.4\textwidth]{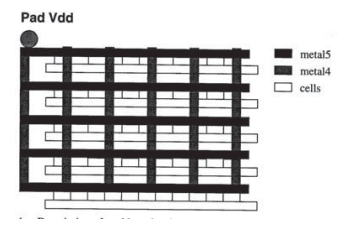}}
\caption{Schematic Power Grid}
\label{fig}
\end{figure}
    \item Add a current source to each cell of the mesh, to compute the magnitude of switching current originating from that cell.
    \item The RC parasitic and switching current information thus obtained is used to calculate dynamic power dissipation for the schematic.
    \item If the Power dissipation numbers are below a pre-defined threshold, we can realize the schematic grid using VDD and VSS stripes.
    
\end{itemize}

    \begin{figure}[htbp]
\centerline{\includegraphics[width=0.5\textwidth]{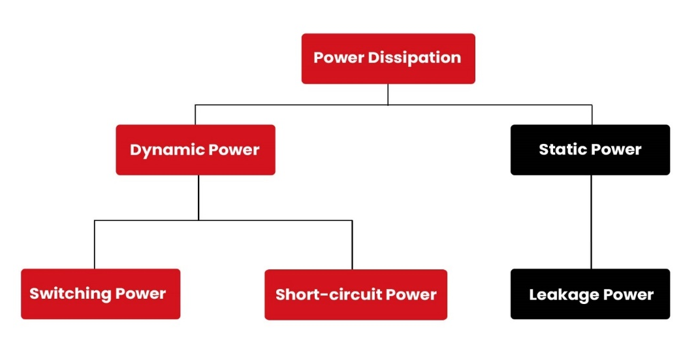}}
\caption{Power Dissipation}
\label{fig}
\end{figure}

Researchers have also tried integrating Power Planning into the Floor Planning stage thereby letting the optimizing algorithms to optimize the floor plan while taking into consideration the parallel creation of an apt power mesh.[23,24] Shuo Zhou et al. have documented the integration of power planning with the Bounded Slicing Grid (BSG) floorplan representation as early as the year 2001. The proposed algorithm was aimed to complete the two planning stages simultaneously on a BSG representation with a view to optimize placement area and power resources.[23] Han Liying et al. in the year 2009 have documented an integrated power and floor planning algortihm based on the Genetic Algorithm (GA) method of optimization.

\subsection{Placement}
Based on the ground-work defined in Floor Plan stage, the objective of Placement stage is to optimize timing and routing within the design. Timing for any block is a function of the cell and net delay between all logically connected paths in the design. Cell delay is a property of the cell. By manipulating the VT flavor of an instance, its cell delay can be changed. Lower the Voltage Threshold (VT), lower the cell delay for the instance. Net delay on the other hand, is a function of the parasitics corresponding to interconnects in the design. These RC Parasitics either be derived from Wire Load Models (WLMs) which are in the form of look-up tables provided by the foundry or from Virtual Route (VR) parameters.
Virtual Route model tends to provide more accurate parasitic extraction than WLM models leading to more refined timing calculations.

Need for placement:

Key factor in determining performance of the circuit: As it indirectly dictates the routing length of wires, it plays a role in determining the delay associated with each wire.

Determines routing ability of the design: A well placed design will have no problems with routing.

Decides distribution of heat on the die surface ; uneven temperature profiles can lead to reliability and timing issues
Power consumption also gets affected by placement

Goal for placement stage:

•	optimize routing resources

•	optimize die utilization

•	reduce routing congestion and hotspots

\subsection{Clock Tree Synthesis}
Once all the core logic has been placed during the previous stage, we start building paths for transmission of clock signal throughout the block. Clock Tree Synthesis is a very critical process owing to the numerous functional failures that may arise out of incorrectly built clock architectures. Misalignment in clock paths may cause setbacks such as failure to meet timing and skews, false outputs, data corruption or even chip-failure. CTS involves building of accurate clock paths and addition of buffers to balance these clock paths by minimizing clock skew. These addition of buffers also aids in reducing hold slack value. At the timing of defining clock tree, signal is derived from source clock for the block and traversed across numerous splits in the tree to ultimately reach the CK pins of flops. If any of these CK endpoint pins are defined to be 'Don't touch' or 'ignore pins', then they are bypassed during clock tree synthesis. Alternatively if any pin is defined to be a 'through pin', it is assumed that the clock path is already built and signal input to the pin is provided.

Types of skews:

•	Global skew achieves zero skew between two synchronous pins without considering logic relationship.

•	Local skew achieves zero skew between two synchronous pins while considering logic relationship.

•	If clock is skewed intentionally to improve setup slack then it is known as useful skew.

.\\Inputs to CTS stage:
\begin{itemize}
    \item Placement Data
    \item Clock Specification File
    \begin{itemize}
    \item Maximum and Minimum insertion delay
    \item Target Skew
    \item Maximum transition value
    \item Non Default Rules (NDR)
    \item Auto CTS root pin
    \item Preferred metal layers for clock
    \item Type of buffers
    \item Latency
    \item Maximum fanout
    \item Maximum capacitance value , etc.
\end{itemize}
\end{itemize}

.\\Steps involved in CTS:
    \begin{itemize}
    \item Synthesize the Clock Tree
    \item Optimize the Clock tree. This is done by
    \begin{itemize}
    \item Buffer relocation
    \item Buffer sizing
    \item Gate relocation
    \item Gate sizing
    \item Improve skew
    \item Delay insertion
\end{itemize}
\item Perform inter-clock balancing
\begin{itemize}
    \item Between 2 flip flop delay balancing has to be done
    \item Clock group between which balancing has to be specified
\end{itemize}
\end{itemize}

While working on lower process nodes, it is observed that with the increase in sequential logic coupled with shorter and more dominant wire delays, it becomes increasingly complex and critical to balance clock skew. While building clock architecture, we generally have a source clock which distributes clock signal throughout the block via clock nets defined using CTS root pins. Clock specifications like max\_capacitance, max\_transition constraints along with max and/or min insertion delays are defined for every root pin and are thus implemented across all clock nets originating from the pin. Guirong Wu et al. in the year 2009, have proposed a more efficient clock splitting methodology to build better clock architectures. The proposed methodology talks about splitting the main source clock into multiple pseudo clock sources at transistor level based on the number of fanouts for every split. This would ultimately help with DRVs as well owing to the split distribution of fanouts, along with more balanced clock skew.[25]

Siong Kiong Teng et al. have proposed another "Regional Clock-Splitting" methodology in their 2010 publication titled "Regional Clock Gate Splitting Algorithm for Clock Tree Synthesis".[26] The said methodology involves the following steps:
    \begin{itemize}
    \item Clock Gate Marking\\Upon placement of standard cells, macros and other physical cells in the layout, all Clock Gating (asynchronous) cells are identified. Each of these Clock Gating (CG) cells are allotted a bounding box that encompasses all direct fanouts related to that cell. This bounding box is demarcated to compare the skews across clock nets affiliated to each of these cells. Simply put, a larger bounding box area implies that the fanouts are located farther away from the CG cell thereby drawing additional interconnect wire length and giving rise to higher skew. 
    \item Clock Gate Splitting\\In the event where any CG cell is associated with any setup violations along with the area encompassed by the corresponding bounding box being greater than a pre-defined threshold, the clock gate cell will undergo splitting. While splitting, the bounding box will split at half the original length in its dominant direction i.e if any bounding box is horizontally oriented, it will split at X/2 distance keeping its Y metric stable, and vice-versa. This splitting is recursively carried out until bounding box area meets the threshold criteria defined earlier. Once such a stage is achieved, the parent clock gate will split 'n' times, where 'n' is the number of times the corresponding bounding box had split in order to meet threshold. The study claims to have achieved reduced post-split insertion delay along with a robust clock architecture with balanced skews.[26]
\end{itemize}

\subsection{Routing}
The Physical Design flow should have completed placement of standard cells, IPs, macros, physical cells and built the entire clock architecture before going for routing. Simply put, routing provides net or wire connectivity between all instances on the layout. Routing creates a grid for transmission of all signals throughout the design. Hence the interconnect nets it creates are also called "signal nets". Similarly CTS stage creates "clock nets" and Power Planning generates "power nets". The aim of routing is to facilitate appropriate connectivity between instances in the design in accordance with the metal stack approved by industry, while also trying to optimize design with respect to possible congestion hotspots. There are two stages within routing methodology:
    \begin{itemize}
    \item Global Routing\\Global Routing, also called Early Routing in some cases is the stage wherein routing metal is allocated to appropriate layers along with channel track alignment. It helps build a general topology for the resources to be used during Detailed Routing.
    \item Detailed Routing\\Detailed Routing, also known as Nano Route in some cases makes use of the foundation laid down by Early routing, to actually route all interconnects in the design. Detailed routing weighs the trade-off between conservation of routing resources and LPA congestions, and optimizes its function accordingly. DFM and congestion checks are carried out for the same purpose.
\end{itemize}

Hao Tang et al. have published a comprehensive survey on the Steiner Tree Algorithm for Global Routing in their 2020 research work titled "A Survey on Steiner Tree Construction and Global Routing for VLSI Design". Steiner Tree is an extension of the Minimum Spanning Tree problem that is commonly used in processing methodologies. The objective of this algorithm is to optimize the routing graph or network to have minimum traversal cost. All vertices in the said graph that constitute the minimum cost tree are referred to as "demand points". All such demand points are connected via horizontal and vertical lines on a mesh referred to as the "underlying graph".[27,28]
    \begin{figure}[htbp]
\centerline{\includegraphics[width=0.4\textwidth]{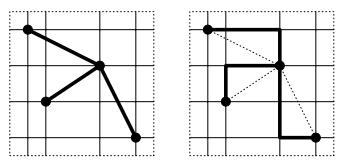}}
\caption{Dominant Points and the Underlying Graph}
\label{fig}
\end{figure}

The 1-Steiner optimization is further implemented on these underlying graphs, wherein a few common points are identified as "1-Steiner Points". These 1-Steiner points is chosen such as to minimize the routing distance (horizontal and vertical) between the dominant points.[28]
    \begin{figure}[htbp]
\centerline{\includegraphics[width=0.4\textwidth]{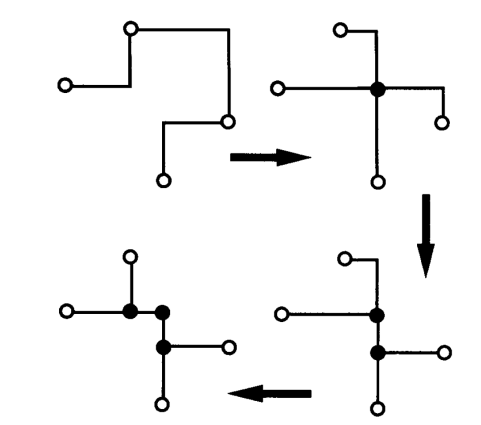}}
\caption{Identification of 1-Steiner points}
\label{fig}
\end{figure}

Just like the Steiner Tree algorithm, other minimum distance algorithms like PRIM, Bounded PRIM, Bounded Radius Spanning Tree have also been implemented to optimize routing for a layout.[28]
.\\Inputs to Routing stage:
\begin{itemize}
    \item Netlist with location of blocks and location of pins (after CTS has been completed)
    \item Timing budget for critical net
    \item Technology File
    \item TLU+ File (Commonly included along with the Technology File .tf)
    \item SDC
\end{itemize}

.\\Checklist before Routing:

\begin{itemize}
    \item Placement completed
    \item CTS completed
    \item Power and ground nets routed
    \item Estimated congestion is acceptable
    \item Estimated Timing – acceptable (~ 0 ns slack )
    \item Estimated max cap/trans – no violations
\end{itemize}

Routing Congestion: When designing chips on lower process nodes and critical utilization factors, it just may happen that routing resources would get crowded in a certain area. Such congestions may have been caused due to lack of routing resources or tracks or simply lack of adequate area to establish the routes. This is a common cause for Density-related DRCs and even EM-IR failures on-chip. 

\subsection{Physical Verification}

Physical verification is a process whereby complete Layout design is verified via EDA software tools to ensure correct logical functionality and manufacturability

Physical Verification involves the following validation checks:

\begin{itemize}
    \item Design Rule Check\\DRCs are of two kinds: Base DRCs and Metal layer DRCs ; of which Base DRCs need to be cleared in the floorplan or placement stage of the flow. These may include placement legalization issues, track misalignment violations, density and utilization requirements not being met, incorrect instance orientation and physical cell violations. Metal layer DRCs on the other hand can be cleared in the Post-Route database as well. These issues range from overlapping vias to metal shorts to even minimum spacing violations between instances and nets.
    \item Layout versus Schematic\\In this process, the final streamout layout GDS is compared against the golden schematic netlist to check for any mismatches or missing instances. All connectivity issues (i.e opens and shorts) are also checked and fixed during cleaning LVS for a design.
    \item Antenna Check
    \item Electrical Rule Check
\end{itemize}

    \begin{figure}[htbp]
\centerline{\includegraphics[width=0.5\textwidth]{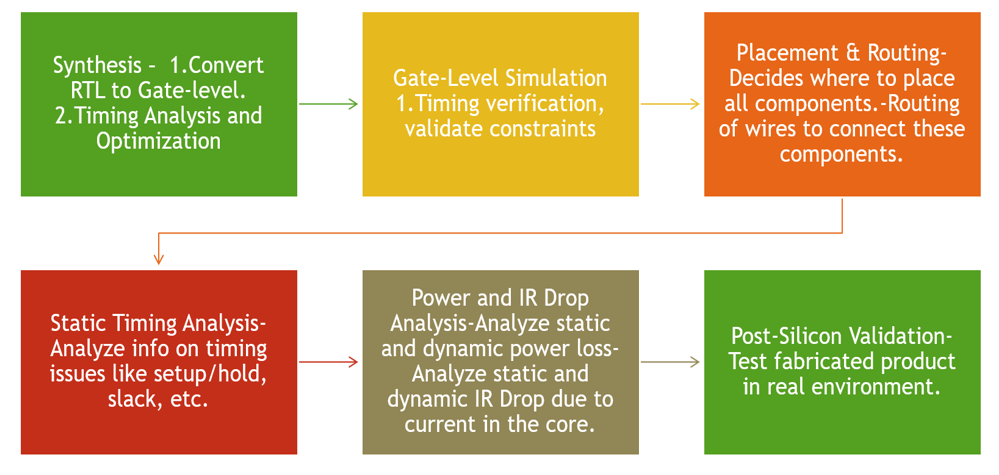}}
\caption{Back End Flow}
\label{fig}
\end{figure}

\section*{Conclusion}
In this paper, we have garnered and compiled a comprehensive study on all the different aspects and methodologies encompassing any robust ASIC Back End Flow. As we move further onto a path that supersedes Moore's Law, it is observed to be increasingly complex to develop even lower process nodes owing to the difficulties in power management, timing closure and conservative routing versus congestion trade-off.[29] While automation aided by EDA tools plays a key role in simplifying and accomplishing design scope, their optimization could be hindered by the aforementioned challenges. Research work compiled by Inamul Hussain et al. captures the gist of these challenges and points towards the use of CNTFETs logic family to design chips at lower technology nodes.[5] However, the ever-developing EDA tools prominently aided by Cadence Design Systems and Synopsys Inc. are able to provide adequate coverage to even use MOSFETs at such lower processes. N.G Aaron et al. have proposed a benchmark for gauging the accuracy and QoR obtained from these EDA tools, which measures the tool performance as well as post-optimzation utility in formulating and displaying results.[8] However, throughout our study it was observed that a well-curated Physcial Design flow is as important as the computational calibre of EDA tools to achieve optimum results and desired functionality.[30,31] 

The input to such a flow is in the form of a synthesized verilog netlist supported by a set of design constraints. Research proposed by Cortadella et al. and Satoshi Ohtake et al. captures the different ways of optimizing the synthesis process with critical path targetting methodologies like High-level Synthesis and Mapping Point Preserving-Logic Synthesis respectively.[12,13] These synthesized .vg, .sdc files along with foundry collaterals like timing libraries, lef files etc are provided as input to the PD flow which is covered in great detail by A. Kahng et al. in their publication titled "VLSI Physical Design: From Graph Partitioning to Timing Closure”.[14] The introductory stage to Physical Design is Floor Planning. Floor planning involves laying a ground-work for the instances to be placed upon. There are a few different ways to represent a floorplan of which the O-Tree and B*Tree are widely regarded as the most efficient. Naushad Manzoor Laskar et al. have compiled a comprehensive survey on all the different Floor Plan representations and algorithms like Simulated Annealing (SA), Genetic Algorithm (GA) and Particle Swarm Optimization (PSO) that are used for optimizing floorplan for the said representations.[18] Researchers have also tried integration of the Power Planning stage into Floor Plan optimization. Once the floorplan is decided upon and the Power/Ground (PG) mesh is generated, the design enters P\&R stage which comprises of Placement, Clock Tree Synthesis (CTS) and Routing. Based  on  the  ground-work  defined  in  Floor  Plan  stage,the  objective  of  Placement  stage  is  to optimize  timing  androuting within the design. In CTS, the clock heirarchies are defined throughout the design for timely and balanced propagation of clock signal. However, balancing clock skews at lower process nodes is a complex yet critical task which even some EDA tools may find challenging to optimize. Siong Kiong Teng et al. have proposed an efficient "Clock Gate Splitting" algorithm to allow a more robust clock architecture with balanced skews and insertion delays. [26] Once the clock heirarchy is established, the design is ready to enter Routing phase wherein interconnects between on-chip instances are established in a two step process: the superficial Global Routing and the more in-depth Detailed Routing. As the main aim of routing phase is to optimize and conserve the routing resources throughout the design, a fair few minimum distance optimization algorithms come to the fray. Of these, the Steiner Tree Algorithm has been studied and documented in detail during the course of this survey.

The design methodologies, optimization algorithms and ASIC flow information garnered and documented in this review may serve as a comprehensive knowledge-point for improving upon the existing methodologies and practices.

\end{document}